\documentclass[10pt,aps,prd,nofootinbib,amsfonts,amssymb,amsmath,superscriptaddress,longbibliography,floatfix]{revtex4-2}
\pdfoutput=1
\usepackage{amsthm}
\usepackage{aas_macros}
\usepackage[utf8]{inputenc}
\usepackage[british]{babel}
\usepackage[Symbolsmallscale]{upgreek}
\usepackage[protrusion=true,tracking=true,kerning=true,spacing=true,final,babel=true]{microtype}
\usepackage{physics}
\usepackage{xcolor}
\usepackage{graphicx}
\usepackage{subfigure}
\usepackage[final]{hyperref}
\usepackage{nth}
\usepackage{cancel}
\usepackage{enumerate}
\usepackage[normalem]{ulem}
\usepackage{xspace}
\usepackage{bbm}
\usepackage{comment}

\graphicspath{{./figures/}
		      }

\def\be{\begin{equation}}
\def\ee{\end{equation}}
\def\bi{\begin{itemize}}
\def\ei{\end{itemize}}
\def\ben{\begin{enumerate}}
\def\een{\end{enumerate}}

\newcommand{\bs}[1]{\boldsymbol{#1}}
\newcommand{\mb}[1]{\mathbf{#1}}

\newcommand{\myfig}[1]{Figure~\ref{#1}}
\newcommand{\mytab}[1]{Table~\ref{#1}}

\defcitealias{abb+2020}{NG12.5}

\begin{document}
\title{Analytic distribution of the optimal cross-correlation 
statistic for stochastic gravitational-wave-background 
searches using pulsar timing arrays}

\author{Jeffrey~S.~Hazboun}
\affiliation{Department of Physics, Oregon State University, Corvallis, OR 97331, USA}
\email{jeffrey.hazboun@oregonstate.edu}

\author{Patrick M. Meyers}
\affiliation{Theoretical Astrophysics Group, California Institute of Technology, Pasadena, California 91125, USA}

\author{Joseph D.~Romano}
\affiliation{Department of Physics and Astronomy, Texas Tech University, 
Lubbock, TX 79409-1051, USA}
\email{joseph.d.romano@ttu.edu}

\author{Xavier Siemens}
\affiliation{Department of Physics, Oregon State University, Corvallis, OR 97331, USA}

\author{Anne M.~Archibald}
\noaffiliation

\date{\today}

\begin{abstract}
We show via both analytical calculation and numerical 
simulation that the optimal cross-correlation statistic (OS) 
for stochastic gravitational-wave-background (GWB) 
searches using data from
pulsar timing arrays follows a {\em generalized chi-squared} (GX2) 
distribution---i.e., a linear combination of chi-squared 
distributions with coefficients given by the eigenvalues of
the quadratic form defining the statistic.
This observation is particularly important for calculating 
the frequentist statistical significance of a possible 
GWB detection, which depends on the exact form of the 
distribution of the OS signal-to-noise ratio (S/N)
$\hat\rho \equiv \hat A_{\rm gw}^2/\sigma_0$
in the absence of GW-induced cross correlations (i.e., the null distribution).
Previous discussions of the OS have incorrectly assumed 
that the analytic null distribution of $\hat\rho$ is 
well-approximated by a zero-mean unit-variance Gaussian 
distribution.
Empiricial calculations show that the null distribution of 
$\hat\rho$ has ``tails" which differ significantly from 
those for a Gaussian distribution, but which follow 
(exactly) a GX2 distribution.
So, a correct analytical assessment of the statistical significance
of a potential detection requires the use of a GX2 distribution.
\end{abstract}
\maketitle

\section{Introduction}
\label{s:intro}

Nanohertz-frequency gravitational waves (GWs) from supermassive binary black holes (SMBBHs) should
be detected in the near future by pulsar timing arrays~\cite{sejr13,Taylor-et-al:2016a,Pol-et-al:2021}.
The North American Nanohertz Observatory for Gravitational Waves
(NANOGrav)~\cite{ransom+19} has already reported strong statistical evidence for a 
common-spectrum red-noise process across pulsars in a search for an
isotropic stochastic gravitational-wave background (GWB) in their 12.5-year dataset~\cite{abb+2020}.
This finding was followed by analyses of Parkes Pulsar Timing Array (PPTA) data~\cite{Goncharov:2021oub},
European Pulsar Timing Array (EPTA) data~\cite{Chen:2021rqp},
as well as the International Pulsar Timing Array (IPTA) second data release~\cite{Antoniadis:2022pcn}.
All of these studies confirmed the presence of a common-spectrum red-noise process across pulsars.
Currently, however, there is little evidence for the quadrupolar 
spatial correlations~\cite{hd83} necessary to make a confident claim of detection of the GWB.
Analytical work and simulations~\cite{sejr13,Taylor-et-al:2016a,Pol-et-al:2021} show that the additional 
statistical evidence in the spatial correlations needed to claim a detection
will come after analyzing larger datasets with more pulsars and longer data spans,
or combining existing datasets from other pulsar timing array collaborations through the IPTA.

A computationally efficient technique used by the pulsar timing community to calculate the significance of the
cross-correlations is the so-called optimal statistic (OS)~\cite{abc+2009,ccs+2015,Vigeland-et-al:2018}.
This statistic is an unbiased estimator for the square of the GWB amplitude $A_{\rm gw}^2$
derived by maximizing the logarithm of the likelihood ratio, and it can be related to the
Bayesian odds ratio between a model with correlations and a model without correlations via the Laplace
approximation~\cite{Romano-Cornish:2017}. The work in this paper concerns the calculation
of the probability distribution for the OS signal-to-noise ratio (S/N)
$\hat\rho\equiv \hat A_{\rm gw}^2/\sigma_0$, where $\sigma_0^2$ is the variance of the estimator 
$\hat A_{\rm gw}^2$ in the absence of GW-induced spatial correlations.

The distribution for $\hat\rho$ in the absence of such 
spatial correlations is called the {\em null distribution}, and 
is denoted by $p(\hat\rho|H_0)$.
$H_0$ is the null hypothesis---i.e., the hypothesis that there 
are no GW-induced spatial correlations in the data. 
Although $H_0$ assumes no spatial correlations, it does 
allow for the presence of a non-zero common-spectrum 
red-noise process as is currently observed, 
whose amplitude $A_{\rm cp}$ is determined from a joint 
noise analysis for the pulsars in the array.
As such, the null distribution $H_0$ depends on the 
particular value of $A_{\rm cp}$---i.e., 
$H_0 = H_0(A_{\rm cp})$. 
However, to simplify the notation in what follows, we will 
not explicitly display the $A_{\rm cp}$ dependence of $H_0$,
although we will investigate the dependence of the null
distribution on the amplitude and spectral shape of the 
common-spectrum red-noise process.

Given the null distribution, we can calculate the 
probability that
our measured S/N, denoted $\hat\rho_{\rm obs}$, 
could have resulted from noise alone.
This is called the $p$-value and is defined as
\be
p
\equiv{\rm Prob}(\hat\rho> \hat\rho_{\rm obs}|H_0)
\equiv\int_{\hat\rho_{\rm obs}}^\infty\>
p(\hat\rho|H_0)\,{\rm d}\hat\rho\,.
\label{e:p}
\ee
The false alarm probability $\alpha$ is similarly defined:
\be
\alpha \equiv{\rm Prob}(\hat\rho> \hat\rho_{\rm th}|H_0)
\equiv\int_{\hat\rho_{\rm th}}^\infty\>p(\hat\rho|H_0)\,{\rm d}\hat\rho\,,
\label{e:alpha}
\ee
where $\hat\rho_{\rm th}$ is the {\em detection threshold}, 
above which one would reject the null hypothesis and 
claim detection of a GWB.
The detection threshold is typically chosen so that the false alarm 
probability has a sufficiently small value, e.g. 
$\alpha < 1\times 10^{-3}$ above that threshold.
Note that the right hand sides of both \eqref{e:p} and \eqref{e:alpha}
can also be written as $1-{\rm CDF}(x|H_0)$, where ${\rm CDF}(x)$ 
is the standard cumulative distribution function and 
$x=\hat\rho_{\rm obs}$ and $\hat\rho_{\rm th}$, respectively.

By construction, both the $p$-value and the false alarm
probability (or, equivalently, the choice of the detection 
threshold) depend on the detailed form of the null distribution. 
In previous work~\cite{abc+2009,ccs+2015,Vigeland-et-al:2018}, 
appealing to the central limit theorem (and also for simplicity), 
this distribution was assumed to be Gaussian.
Here, we show that the null distribution for the OS follows
a {\em generalized chi-squared} (GX2) distribution~\cite{dg2021}
and we explore the consequences of this observation.
(A paper by Cordes and Shannon~\cite{cordes-shannon:2012} also
noted that the distribution of the cross-correlation statistic 
was highly non-Gaussian and skewed, but did not identify it as a GX2 
distribution.)
We show that differences between the GX2 and Gaussian distributions 
can be significant for current pulsar timing array configurations 
(defined by numbers of pulsars, observation spans, noise parameters, etc.), 
as well as the amplitude and spectral shape of the 
common-spectrum process, especially in the tails of the distribution.
Thus, the GX2 distribution should be used to calculate more accurate 
$p$-values in the case of a GW detection. 
In particular, we show that the Gaussian distribution
assumption for the null distribution of the OS leads to overestimates
of the significance of a potential detection, i.e., smaller $p$-values, than for
the GX2 distribution. 

We also compare the GX2 and Gaussian distributions of the OS to the null 
distributions obtained from doing phase-shifts~\cite{Taylor-et-al:2016b}
and sky scrambles~\cite{Cornish-Sampson:2016} of the PTA data.
The phase-shifts are analogous to applying time-shifts in ground-based 
GW detectors, while sky scrambles replace the actual pulsar locations
with random sky locations, thus breaking all spatial correlations except 
for the direction-independent monopole.
Both of these techniques work well at removing GW-induced correlations,
and are now standard methods to determine null distributions for our statistics, 
be it the OS or the Bayes factor, directly from our data~\cite{abb+2020}.  
We show that the GX2 distribution is an excellent fit to the phase-shifted
/ sky-scrambled data compared with the standard
Gaussian approximation, especially in the tails of the distribution (where it matters most)
at larger values of the OS signal-to-noise ratio $\hat\rho$.

The rest of the paper is organized as follows:
In Sec.~\ref{s:derivation}, we summarize the results of \cite{dg2021},
explaining how GX2 distributions arise whenever one has a (symmetric)
quadratic combination of random variables satisfying a 
multivariate Gaussian distribution.
We then show, in Sec.~\ref{s:application}, that the OS used for 
PTA searches is an example of such a quadratic combination, explicitly 
describing the various pieces that enter the calculation of the 
eigenvalues needed for the GX2 distribution.
In Section~\ref{s:results}, we first calculate GX2 distributions for the OS for 
several different sets of simulated data, in the absence 
of simulated GW-induced cross correlations.
We show how these distributions depend on the relative contribution 
of red and white noise, the number of pulsars, fitting to a timing
a model, etc., and compare these distributions with unit (i.e., 
standard normal) Gaussians, 
which often deviate significantly from the GX2 distributions in the 
tails of the distributions.
We then compare the GX2 
distribution for the OS with the null distribution 
of $\hat\rho$ obtained by phase-shifting the NANOGrav
12.5-yr data~\cite{abb+2020}.
Finally, we conclude in Sec~\ref{s:discussion} by discussing 
possible extensions / modifications to the calculations 
presented here---e.g., which might simplify the calculation 
of GX2 distributions for realistic PTA datasets.
Appendix~\ref{s:empirical_tail_fitting} describes a simple
``tail-fitting" approach that allows us to extrapolate the 
tail of the OS null distribution beyond empirically determined 
phase-shift or sky-scramble values.

\section{Mathematical formalism}
\label{s:derivation}

By definition, the \emph{generalized chi-square} (GX2)
distribution is the probability distribution of 
a quadratic form 
of multivariate-Gaussian random variables 
$\mb{x}\sim {\cal N}(\bs\mu,\bs\Sigma)$:
\be
q(\mb{x}) \equiv \frac{1}{2}\mb{x}^T \mb{Q}_2 \mb{x} 
+ \mb{q}_1^T\mb{x} + q_0\,,
\ee
where
\be
p(\mb{x}|\bs\mu,\bs\Sigma)
= \frac{1}{\sqrt{{\rm det}(2\pi\bs\Sigma})}
e^{-\frac{1}{2} (\mb{x}-\bs\mu)^T \bs\Sigma^{-1}(\mb{x}-\bs\mu)}\,.
\ee
Here $q_0$ is a real constant, $\mb{q}_1$ is a real vector of the
same dimension as $\mb{x}$, 
and $\mb{Q}_2$ is a real symmetric matrix $\mb{Q}_2^T=\mb{Q}_2$.
As we shall describe in Sec.~\ref{s:application}, 
the cross-correlation statistic that we are interested in 
has the simpler form
\be
q(\mb{x}) \equiv \frac{1}{2}\mb{x}^T \mb{Q} \mb{x}\,,
\qquad
\mb{x}\sim {\cal N}(\mb{0},\bs{\Sigma})\,.
\label{e:q(x)}
\ee
In this last equation we have dropped 
the subscript 2 from $\mb{Q}_2$ to simplify the 
notation,
since there is no potential to confuse it with a
linear or constant term.
We will work with this simpler form for the rest of the paper.

To use the formalism of \cite{dg2021} to explicitly compute
the GX2 distribution of $q(\mb{x})$, we need to write
\eqref{e:q(x)} as a linear superposition of standard normal
distributions $\mb{v}\sim {\cal N}(\mb{0},\mb{1})$.
This is done via a series of eigenvalue/eigenvector
decompositions which we summarize below:

Following \cite{dg2021}, we start by converting 
$\mb{x}\sim{\cal N}(\mb{0},\bs{\Sigma})$ 
to a vector of uncorrelated standard normal distributions
$\mb{z}\sim {\cal N}(\mb{0},\mb{1})$ by finding the eigenvalues 
and eigenvectors of $\bs{\Sigma}$:
\be
\mb{D} = \mb{E}^T \bs{\Sigma}\mb{E}
\quad{\rm or}\quad
\bs{\Sigma} = \mb{E} \mb{D}\mb{E}^T\,,
\label{e:D}
\ee
where $\mb{D}$ is a diagonal matrix of eigenvalues
$\mb{D} = {\rm diag}(\sigma^2_{1}, \sigma^2_{2}, \cdots)$
and $\mb{E}$ is an orthogonal matrix 
(i.e., $\mb{E}^T=\mb{E}^{-1}$)
whose columns are the corresponding (orthonormal) 
eigenvectors of $\bs{\Sigma}$.
Since $\bs{\Sigma}$ is a covariance matrix, we are
guaranteed that its eigenvalues are all positive, hence
the form $\sigma^2_1$, $\sigma^2_2$, $\ldots$.
We then take the square root of $\mb{D}$:
\be
\bs{\Lambda} \equiv \sqrt{\mb{D}}\,,
\label{e:Lambda}
\ee
for which
\be
\mb{x}^T\bs{\Sigma}^{-1}\mb{x}
=\mb{x}^T\mb{E}\mb{D}^{-1}\mb{E}^T \mb{x}
=\mb{x}^T \mb{E}\bs{\Lambda}^{-1}\bs{\Lambda}^{-1}\mb{E}^T\mb{x}
=\mb{z}^T \mb{1} \mb{z}\,,
\ee
where
\be
\mb{z} \equiv \bs{\Lambda}^{-1}\mb{E}^T\mb{x}
\quad\Leftrightarrow\quad
\mb{x} = \mb{E}\bs{\Lambda}\mb{z}\,.
\ee
In terms of $\mb{z}$ the quadratic combination \eqref{e:q(x)} 
has the form
\be
q(\mb{x})
=\frac{1}{2}\mb{z}^T\tilde{\mb{Q}}\mb{z}\,,
\qquad
\mb{z}\sim{\cal N}(\mb{0},\mb{1})\,,
\ee
where
\be
\tilde{\mb{Q}} \equiv 
\bs{\Lambda}^T \mb{E}^T\mb{Q}\mb{E}\bs{\Lambda}\,.
\label{e:Qtilde}
\ee

The final step is to diagonalize $\tilde{\mb{Q}}$,
by finding its eigenvalues $(\tilde e_1, \tilde e_2, \cdots)$ and 
the orthogonal matrix of eigenvectors $\mb{U}$.
This gives
\be
\tilde{\mb{Q}} = \mb{U} {\rm diag}(\tilde e_1, \tilde e_2, ...) \mb{U}^T\,,
\ee
for which
\be
q(\mb{x}) = \frac{1}{2}\sum_{i} \tilde e_i v_i^2\,,
\quad{\rm where}\quad
\mb{v}\equiv \mb{U}^T \mb{z} \sim {\cal N}(\mb{0},\mb{1})\,.
\label{e:sum_over_evals}
\ee
(Note that $\mb{v}$ is standard normal since $\mb{U}$ is an 
orthogonal matrix and $\mb{z}$ is standard normal.)
Since the probability distribution of the square of a standard 
normal distribution is chi-squared distributed with 1 degree of freedom
(DOF), it follows
that $q(\mb{x})$ is a general linear combination of $\chi^2_1$ distributions,
which is the form of the generalized chi-squared distribution discussed in
Ref.~\cite{dg2021}.
The analytic form of this distribution for the quadratic form \eqref{e:q(x)} 
is completely specified by the eigenvalues of 
$\tilde{\mb{Q}}$ defined by \eqref{e:Qtilde}, \eqref{e:D}, and \eqref{e:Lambda}.

Note that the mean and variance of $q(\mb{x})$ can be 
simply written in terms of sums (and sums of squares) of the eigenvalues:
used to construct the optimal statistic.
\be
\mu_q\equiv
\langle q\rangle = \frac{1}{2} \sum_i\tilde e_i
\qquad{\rm and}\qquad
\sigma^2_q\equiv 
\langle q^2\rangle - \langle q\rangle^2 
= \frac{1}{2} \sum_i\tilde e_i^2\,,
\label{e:eval_props}
\ee
where $\langle\ \rangle$ denotes expectation value.
The above results follow from the expansion \eqref{e:sum_over_evals}
with the $v_i^2\sim \chi^2_1$ being statistically independent of 
one another, and each having
${\rm mean}=1$ and ${\rm variance}=2$.

\section{Application to the optimal statistic for PTA searches for GWBs}
\label{s:application}

Now we show that different forms of the OS used for PTA searches 
for GWBs are examples of GX2 distributions.
For more details regarding the OS, we refer the reader to 
see~\cite{abc+2009,ccs+2015,Vigeland-et-al:2018}.

\subsection{Optimal statistic S/N and GWB amplitude estimator}
\label{s:rhohat_Agw2}

The optimal statistic S/N for PTA searches for GWBs
is typically written as 
\be
\hat\rho\equiv
{\hat A^2_{\rm gw}}/{\sigma_0}\,,
\label{e:rhohat}
\ee
where
\begin{align}
&\hat A^2_{\rm gw} 
\equiv {\cal N}\sum_{a<b}\mb{r}^T_a 
\mb{P}_a^{-1}\tilde{\mb{S}}_{ab}\mb{P}_b^{-1}\mb{r}_b\,,
\label{e:A2hat}
\\
&\sigma_0 \equiv {\cal N}^{1/2}\,,
\label{e:sigma0}
\\
&{\cal N}
\equiv \left(\sum_{a<b} {\rm tr}\left[\mb{P}_a^{-1}\tilde{\mb{S}}_{ab}
\mb{P}_b^{-1}\tilde{\mb{S}}_{ba}\right]\right)^{-1}\,.
\label{e:calN}
\end{align}
In the above expressions,
$\hat A_{\rm gw}^2$ is an estimator of the (squared) amplitude 
of the GW signal, $\sigma_0^2$ is its variance in the absence of 
GW-induced spatial correlations, and $\cal N$ is a normalization 
factor constructed from terms involving the total auto-correlated 
power and cross-correlated power in pulsars labeled by 
$a$, $b$ (more about these expressions below).
More compactly,
\be
\hat\rho 
\equiv 
\sum_{a<b}\mb{r}_a^T \mb{Q}_{ab}\mb{r}_b
= \frac{1}{2}\sum_{a,b}\mb{r}_a^T \mb{Q}_{ab}\mb{r}_b\,,
\label{e:rhohat_doublesum}
\ee
where
\be
\mb{Q}_{aa}\equiv\mb{0}\,,
\qquad
\mb{Q}_{ab}\equiv
{\cal N}^{1/2}\mb{P}_a^{-1}\tilde{\mb{S}}_{ab}\mb{P}_b^{-1}
\label{e:Qtilde_OS}
\ee
define the (symmetric) quadratic form for $\hat\rho$.
Note that
summations denoted by $\sum_{a<b}$ run over {\em distinct} pulsar
pairs, while $\sum_{a,b}\equiv \sum_a\sum_b$ double counts the
pulsar pairs (hence the factor of 1/2 in \eqref{e:rhohat_doublesum})
and also includes the auto-correlations (which don't contribute 
to $\hat\rho$ since $\mb{Q}_{aa}=\mb{0}$).

The data $\mb{r}$ are zero-mean multivariate Gaussian random variables
defined by
\be
p(\mb{r}|\vec\theta) = \frac{1}{\sqrt{{\rm det}(2\pi\bs{\Sigma})}}
\exp\left(-\frac{1}{2}\mb{r}^T\bs{\Sigma}^{-1}\mb{r}\right)\,,
\ee
where
\be
\bs{\Sigma} \equiv \langle\mb{r}\mb{r}^T\rangle=
\begin{pmatrix}
\mb{P}_1 & \mb{S}_{12} & \cdots & \mb{S}_{1M} \\
\mb{S}_{21} & \mb{P}_2 & \cdots & \mb{S}_{2M} \\
\vdots & \vdots & \ddots & \vdots \\
\mb{S}_{M1} & \mb{S}_{M2} & \cdots & \mb{P}_M \\ 
\end{pmatrix}
\label{e:Sigma_OS}
\ee
and
\be
\mb{P}_a \equiv \mb{G}_a^T \mb{N}_a \mb{G}_a\,,
\qquad
\mb{S}_{ab} \equiv \mb{G}_a^T \mb{X}_{ab} \mb{G}_b\,,
\qquad a, b =1,2,\cdots, M\,.
\ee
Here, $M$ denotes the number of pulsars, and 
$\mb{G}_a$ is the $G$-matrix for pulsar $a$ \cite{vanHaasteren-levin:2013}, which encodes
information about the timing-model fit
\be
\mb{r}_a = \mb{G}^T_a\,\bs{\delta}\mb{t}_a\,,
\qquad a=1,2,\cdots, M\,,
\label{e:r_Gmatrix}
\ee
where $\bs{\delta}\mb{t}_a$ are the timing residuals for
pulsar $a$.
If we denote the number of TOAs for pulsar $a$ 
by $N_{{\rm TOA},a}$ and the number of timing model 
parameters by $N_{{\rm par},a}$, 
then $\mb{G}_a$ has dimensions 
$N_{{\rm TOA},a}\times(N_{{\rm TOA},a}-N_{{\rm par},a})$,
and $\mb{r}_a$ is a vector with components
\be
[\mb{r}_a]_{\alpha_a}\,,
\qquad
\alpha_a = 1, 2, \cdots, N_a \equiv (N_{{\rm TOA},a}-N_{{\rm par},a})\,.
\ee
The covariance matrix $\bs{\Sigma}$ is thus a symmetric block matrix
and has overall dimension
$(N_1 + N_2 + \cdots + N_M)\times(N_1 + N_2 + \cdots + N_M)$.

The diagonal terms of the covariance matrix involve
the autocorrelations
\be
\mb{N}_a \equiv \langle\bs{\delta}\mb{t}_a\bs{\delta}\mb{t}^T_a\rangle=
\int_0^{f_{\rm Nyq}} {\rm d}f\>
\cos[2\pi f\bs{\tau}_{aa}]\,{\cal P}_a(f) 
+{\cal F}_a \mb{W}_a + {\cal Q}^2_a\mb{1}\,,
\label{e:Na}
\ee
where the last two terms specify the white noise contributions,
and 
\be
{\cal P}_a(f) \equiv {\cal P}_a^{\rm red}(f) + {\cal P}_{\rm gw}(f)
\ee
consists of both intrinsic pulsar red noise 
and a potential common-spectrum red-noise process contribution most 
likely from the GWB.
We assume that both of these red noise contributions can be described
by power-law spectra
\be
{\cal P}_a^{\rm red}(f) \equiv
\frac{A^2_a}{12\pi^2 f^3} \left(\frac{f}{f_{\rm ref}}\right)^{2\alpha_a}\,,
\qquad
{\cal P}_{\rm gw}(f) \equiv
\frac{A^2_{\rm gw}}{12\pi^2 f^3} \left(\frac{f}{f_{\rm ref}}\right)^{2\alpha_{\rm gw}}\,.
\ee
For the GWB formed from the superposition of signals from 
inspiraling super-massive binary black holes in the centers of
merging galaxies, $\alpha_{\rm gw}=-2/3$. 
Finally, $\bs{\tau}_{aa}$ is the time-lag matrix, whose 
components are given by $[\bs{\tau}_{aa}]_{ij}\equiv t_{i_a}-t_{j_a}$, 
which are the differences of the 
TOAs of the pulses from pulsar $a$.

The off-diagonal terms in the covariance matrix are
assumed to have only a GWB contribution
\be
\mb{X}_{ab} \equiv
\langle\bs{\delta}\mb{t}_a\bs{\delta}\mb{t}^T_b\rangle=
\Gamma_{ab}
\int_0^{f_{\rm Nyq}} {\rm d}f\>
\cos[2\pi f\bs{\tau}_{ab}]\,{\cal P}_{\rm gw}(f)\,,
\ee
where 
\be
\Gamma_{ab}
\equiv
\frac{1}{2} 
+\frac{3}{2}\left(\frac{1-\cos\xi_{ab}}{2}\right)
\left[\ln\left(\frac{1-\cos\xi_{ab}}{2}\right)-\frac{1}{6}\right]
+\frac{1}{2}\delta_{ab}
\ee
are the values of the Hellings-and-Downs function, 
$\Gamma_{ab} \equiv \Gamma(\xi_{ab})$, evaluated 
for two pulsars $a$ and $b$ separated by the angle $\xi_{ab}$,
see~\cite{hd83}.
The quantity 
\be
\tilde{\mb{S}}_{ab}\equiv \mb{G}^T_a\tilde{\mb{X}}_{ab}\mb{G}_b\,,
\ee
which enters the expression for the
quadratic form $\mb{Q}_{ab}$, is a normalized version of 
$\mb{S}_{ab}$ defined in terms of
\be
\tilde{\mb{X}}_{ab}\equiv \mb{X}_{ab}/A_{\rm gw}^2\,.
\ee
Note that these cross-correlations are proportional to the {\em spectral shape} 
of the GWB---i.e., they do not depend on its amplitude.

Finally, it is a simple matter to show that the GWB amplitude 
estimator $\hat A^2_{\rm gw}$ can
also be written as a (symmetric) quadratic combination of the 
multivariate Gaussian random variables $\mb{r}$ with
quadratic form
\be
\mb{K}_{ab} \equiv {\cal N}^{1/2} \mb{Q}_{ab}\,.
\ee
So, according to the discussion in Sec.~\ref{s:derivation}, 
both $\hat\rho$ and $\hat A^2_{\rm gw}$ 
will be described by GX2 distributions.
For calculating the distributions of these statistics
in the absence of GW-induced spatial correlations
(i.e., {\em null} distributions), we should set 
$\mb{X}_{ab}=\mb{0}$ in the definition of
$\bs{\Sigma}$ and replace the GW contribution
${\cal P}_{\rm gw}(f)$ to $\mb{N}_a$ by a potential
common-spectrum red-noise process ${\cal P}_{\rm cp}(f)$ 
(with amplitude $A_{\rm cp}$ and spectral index $\alpha_{\rm cp}$), 
which is common to all pulsars.
(The normalized cross-correlation terms $\tilde{\mb{S}}_{ab}$
in $\mb{Q}_{ab}$ do not change since they involve the 
{\em normalized} cross-correlations $\tilde{\mb{X}}_{ab}$.)
These null distributions are obviously also described by 
GX2 distributions since they are special cases of the 
non-null quadratic combinations.

Using \eqref{e:eval_props}, it follows the 
eigenvalues $\tilde e_i$ that specify the GX2
distribution for $\hat\rho\equiv A^2_{\rm gw}/\sigma_0$ 
in the absence of GW-induced
spatial correlations satisfy:
\be
\sum_i\tilde e_i=0
\qquad{\rm and}\qquad
\sum_i\tilde e_i^2=2\,.
\ee
These are consequences of $\hat\rho$ having zero mean
and unit variance for the null distribution case.

\subsection{Optimal statistic pulsar-pair cross-correlation estimators}
\label{s:rho_ab}

One can also construct cross-correlation estimators for individual
pulsar pairs:
\be
\hat\rho_{ab} 
\equiv {\cal N}_{ab}\,
\mb{r}^T_a \mb{P}_a^{-1}\bar{\mb{S}}_{ab}\mb{P}_b^{-1}\mb{r}_b
\equiv
\mb{r}^T_a \bar{\mb{Q}}_{ab}\mb{r}_b\,,
\qquad a<b=1,2,\cdots, M\,,
\label{e:rhoab_hat}
\ee
where
\begin{align}
&\bar{\mb{Q}}_{ab} \equiv  {\cal N}_{ab}\,
\mb{P}_a^{-1}\bar{\mb{S}}_{ab}\mb{P}_b^{-1}\,,
\\
&{\cal N}_{ab}\equiv 
\left({\rm tr}\left[\mb{P}_a^{-1}\bar{\mb{S}}_{ab}
\mb{P}_b^{-1}\bar{\mb{S}}_{ba}\right]\right)^{-1}\,,
\\
&\bar{\mb{S}}_{ab} \equiv \mb{G}^T_a \bar{\mb{X}}_{ab}\mb{G}_b\,,
\\
&\bar{\mb{X}}_{ab} \equiv \mb{X}_{ab}/(\Gamma_{ab} A_{\rm gw}^2)
= \int_0^{f_{\rm Nyq}} {\rm d}f\>
\cos[2\pi f\bs{\tau}_{ab}]\,\bar{{\cal P}}_{\rm gw}(f)\,,
\\
&\bar{{\cal P}}_{\rm gw}(f)\equiv
\frac{1}{12\pi^2 f^3} \left(\frac{f}{f_{\rm ref}}\right)^{2\alpha_{\rm gw}}\,.
\end{align}
Using
\be
\langle \mb{r}_a\mb{r}_b^T\rangle
= \mb{S}_{ab}\,,
\ee
it follows that $\hat\rho_{ab}$ is an unbiased estimator
of the cross-correlated power in the GWB---i.e.,
\be
\langle\hat\rho_{ab}\rangle =\Gamma_{ab} A_{\rm gw}^2\,,
\ee
with variance
\be
\sigma^2_{ab} = {\cal N}_{ab}
+ (A_{\rm gw}^2)^2{\cal N}_{ab}^2\,
{\rm tr}\,[\mb{P}_a^{-1}\bar{\mb{S}}_{ab}\mb{P}_b^{-1}\tilde{\mb{S}}_{ba}
\mb{P}_a^{-1}\bar{\mb{S}}_{ab}\mb{P}_b^{-1}\tilde{\mb{S}}_{ba}]\,.
\ee
In the absence of GW-induced spatial correlations, 
the variance simplifies to 
\be
\sigma^2_{ab} ={\cal N}_{ab} \equiv \sigma^2_{0,ab}\,.
\label{e:sigma^2_0ab}
\ee
Recall that $\Gamma_{ab}$ are the values of the 
Hellings-and-Downs function evaluated for different 
angular separations between the two pulsars 
labeled by $a$ and $b$.
The matrix $\bar{\mb{X}}_{ab}$ is the time-domain
representation of the {\em spectral shape} of 
the GW power spectrum, which (by its definition) 
is independent of the GWB amplitude $A_{\rm gw}$ 
and the spatial correlation coefficients $\Gamma_{ab}$.
It depends on the pulsar pair $ab$ only via the 
discrete times $t_{i_a}$, $t_{j_b}$ of the
timing residuals $\bs{\delta}\mb{t}_a$, $\bs{\delta}\mb{t}_b$,
which enter the time-lag matrix $\bs{\tau}_{ab}$.
This means that $\bar{\mb{X}}_{ab}$ is a rectangular 
matrix with 
dimensions $N_{{\rm TOA},a}\times N_{{\rm TOA},b}$.

Since $\hat\rho_{ab} = \hat\rho_{ba}$, we can write
\be
\hat\rho_{ab} = \frac{1}{2}(\hat\rho_{ab} + \hat\rho_{ba})
=\frac{1}{2}\left(
\mb{r}^T_a \bar{\mb{Q}}_{ab}\mb{r}_b +
\mb{r}^T_b \bar{\mb{Q}}_{ba}\mb{r}_a \right)
=\frac{1}{2} \mb{x}^T \mb{Q} \mb{x}\,,
\ee
where
\be
\mb{x}^T \equiv
\begin{bmatrix}
\mb{r}^T_a
&
\mb{r}^T_b
\end{bmatrix}\,,
\qquad
\mb{Q} \equiv
\begin{bmatrix}
\mb{0} & \bar{\mb{Q}}_{ab}
\\
\bar{\mb{Q}}_{ba} & \mb{0}
\end{bmatrix}\,,
\qquad
\mb{x} \equiv
\begin{bmatrix}
\mb{r}_a
\\
\mb{r}_b
\end{bmatrix}\,.
\ee
Note that $\mb{Q}$ is a symmetric matrix since
$\bar{\mb{Q}}_{ab}^T = \bar{\mb{Q}}_{ba}$.
Note also that $\mb{x} \sim {\cal N}(\mb{0},\bs{\Sigma)}$,
where
\be
\bs{\Sigma}\equiv 
\langle\mb{x}\mb{x}^T\rangle
=\begin{bmatrix}
\mb{P}_a & \mb{S}_{ab} 
\\
\mb{S}_{ba} & \mb{P}_b
\end{bmatrix}\,.
\ee
So given that $\hat\rho_{ab}$ is a symmetric quadratic
combination of multivariate Gaussian random
variables, it obeys a GX2 distribution.

If desired, one can express both the OS signal-to-noise ratio 
$\hat\rho$ and the GWB amplitude estimator $\hat A_{\rm gw}^2$ 
in terms of the pulsar-pair cross-correlation
estimators $\hat\rho_{ab}$:
\be
\hat\rho = 
\frac{\sum_{a<b} \Gamma_{ab}\hat\rho_{ab}/\sigma^2_{0,ab}}
{\sqrt{\sum_{c<d} \Gamma_{cd}^2 / \sigma^2_{0,cd}}}
\qquad{\rm and}\qquad
\hat A^2_{\rm gw} = 
\frac{\sum_{a<b} \Gamma_{ab}\hat\rho_{ab}/\sigma^2_{0,ab}}
{\sum_{c<d} \Gamma_{cd}^2 / \sigma^2_{0,cd}}\,.
\ee
These results are a consequence of 
$\tilde{\mb{S}}_{ab}=\Gamma_{ab}\bar{\mb{S}}_{ab}$ and
the definitions \eqref{e:rhohat}, \eqref{e:A2hat}, and 
\eqref{e:rhoab_hat} 
of $\hat\rho$, $\hat A^2_{\rm gw}$, and $\hat\rho_{ab}$. 
Thus, $\hat\rho$ and $\hat A_{\rm gw}^2$ are simple
noise-weighted and $\Gamma_{ab}$-matched linear
combinations of the $\hat\rho_{ab}$.

\section{Demonstration of the GX2 distribution on PTA data}
\label{s:results}

The exact form of the GX2 distribution varies with the amplitude of the GW
signal and the noise properties of the data. The characteristics of PTA noise
have been studied extensively in the literature
\cite{lam+18optfreq,lam18_optcad,Hazboun-et-al:2019}. The considerations for 
PTA detector characterization vary widely from radiometer noise at the telescope
receiver, to possible intrinsic spin instabilities of the neutron star. However, they can
be split into a few categories including white noise (WN), red noise
(RN), the fit of the deterministic timing model for the radio pulse
times-of-arrival, and the number and sky position of the pulsars. In this
section we demonstrate the influence that these characteristics of PTAs have on
the GX2 distribution.

\begin{figure}[h!tbp]
\centering
\subfigure[]
{\includegraphics[width=0.48\textwidth]{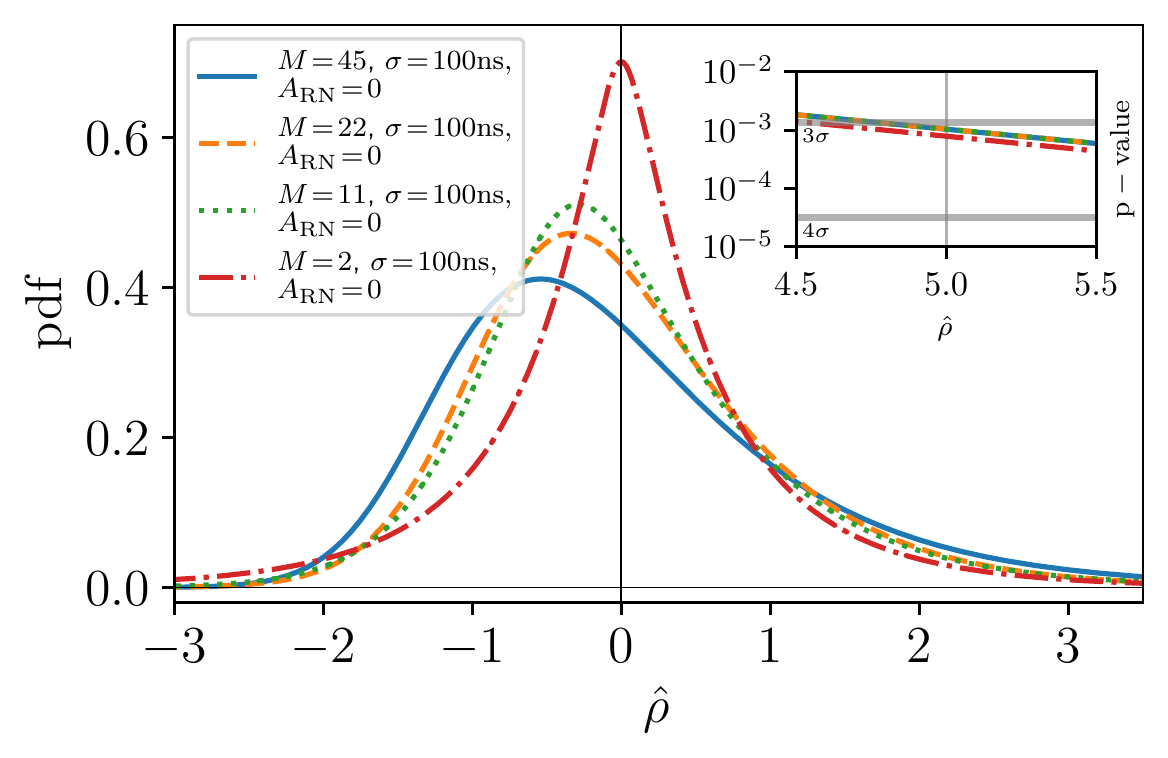}}
\subfigure[]
{\includegraphics[width=0.48\textwidth]{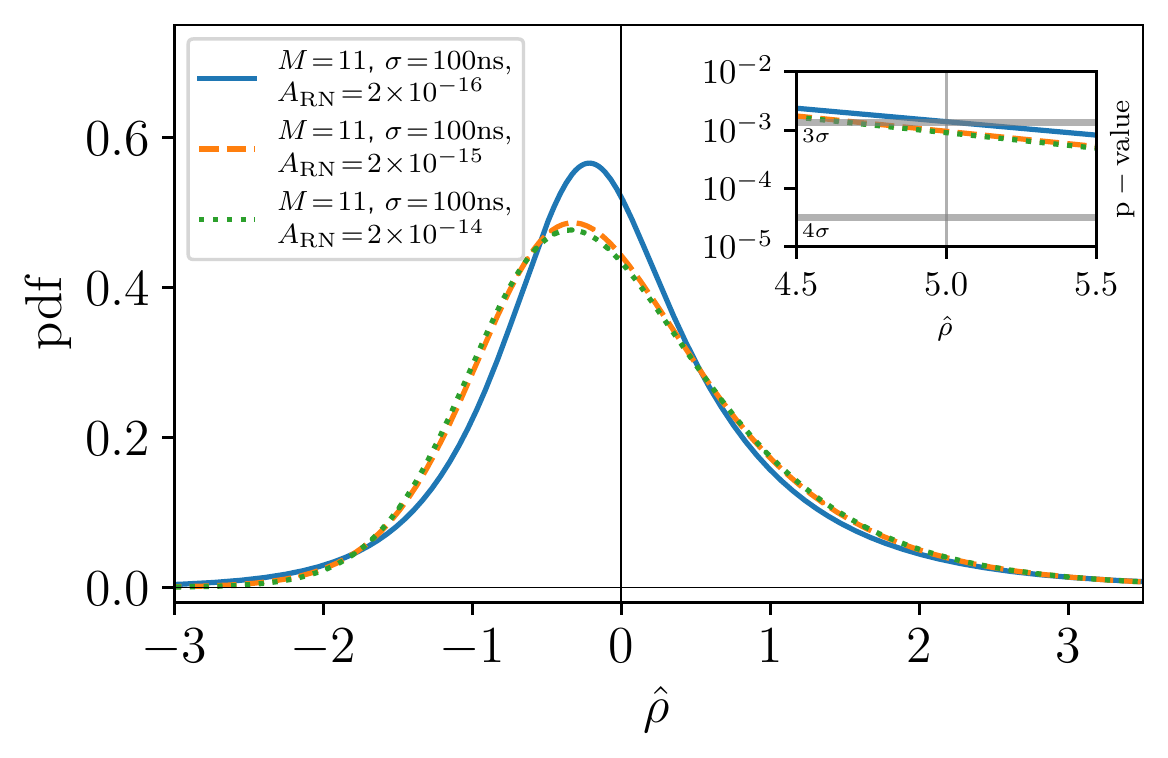}}
\subfigure[]
{\includegraphics[width=0.48\textwidth]{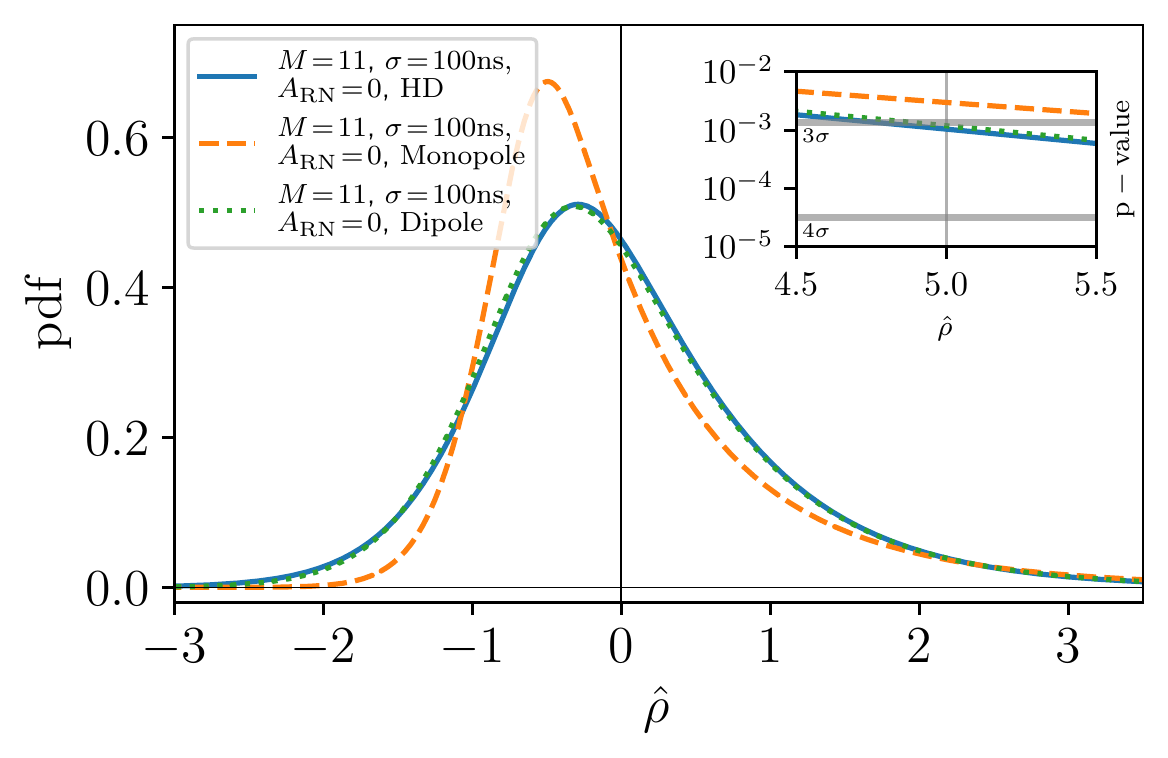}}
\subfigure[]
{\includegraphics[width=0.48\textwidth]{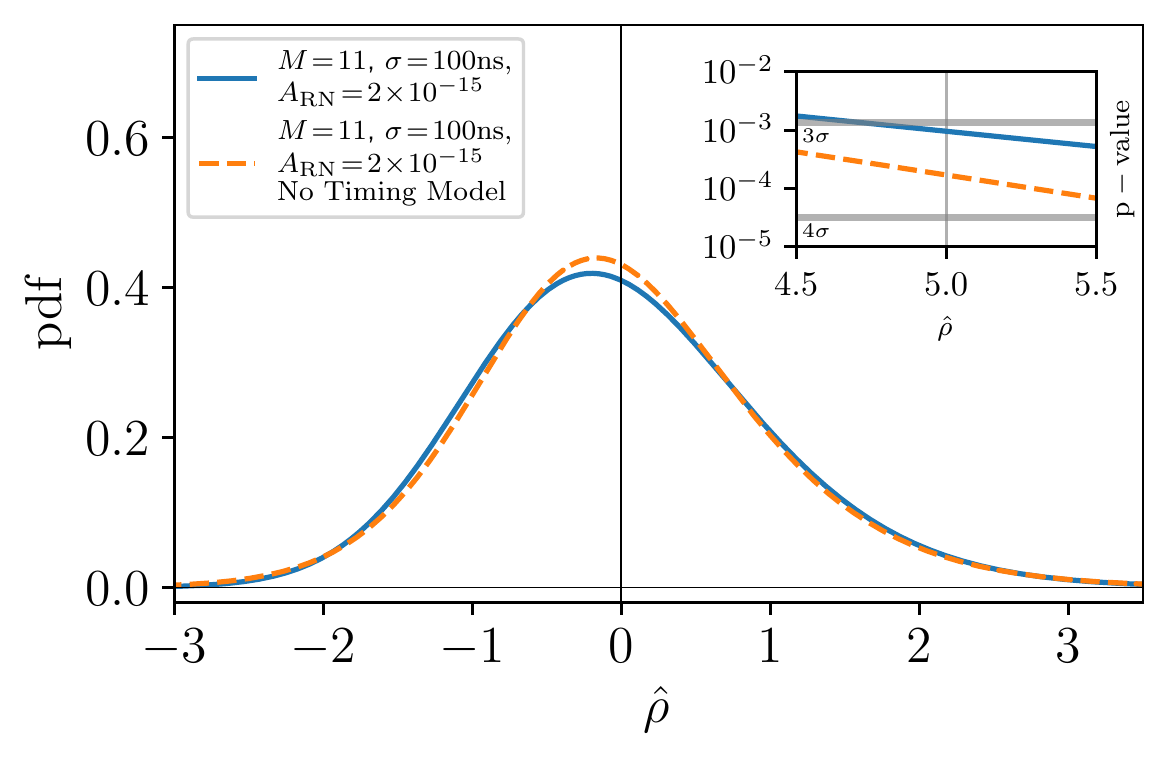}}
\subfigure[]
{\includegraphics[width=0.48\textwidth]{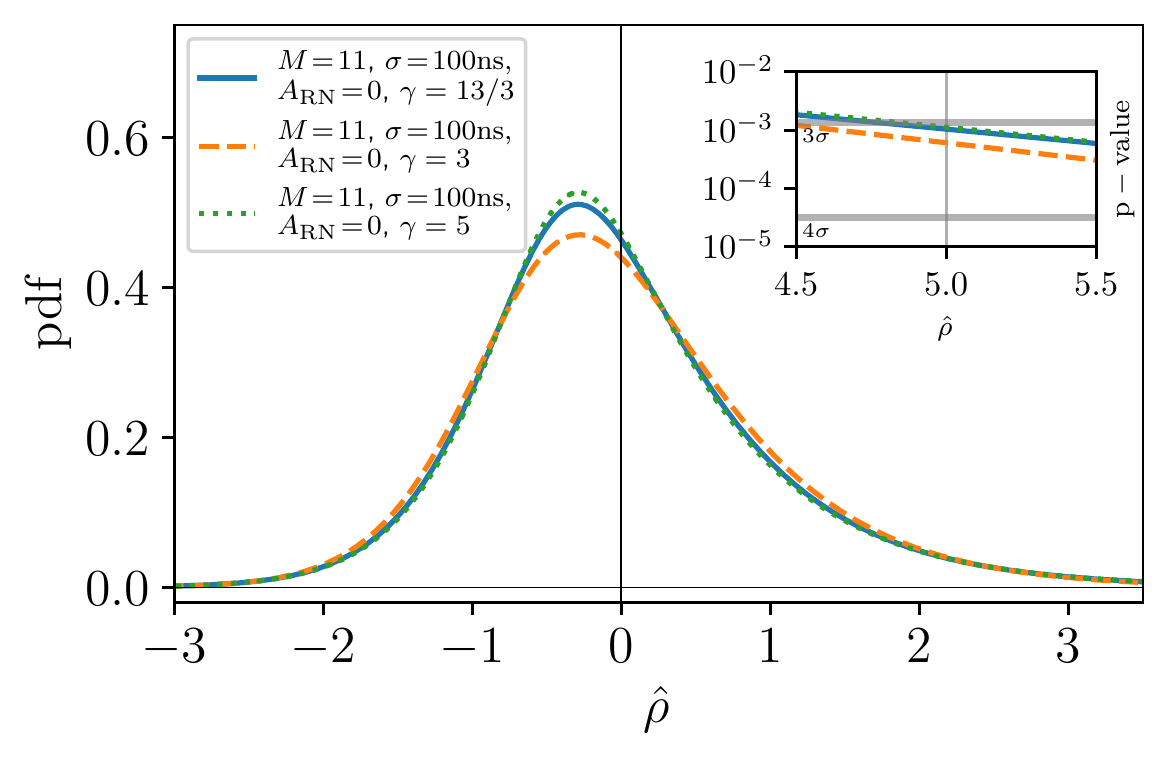}}
\subfigure[]
{\includegraphics[width=0.48\textwidth]{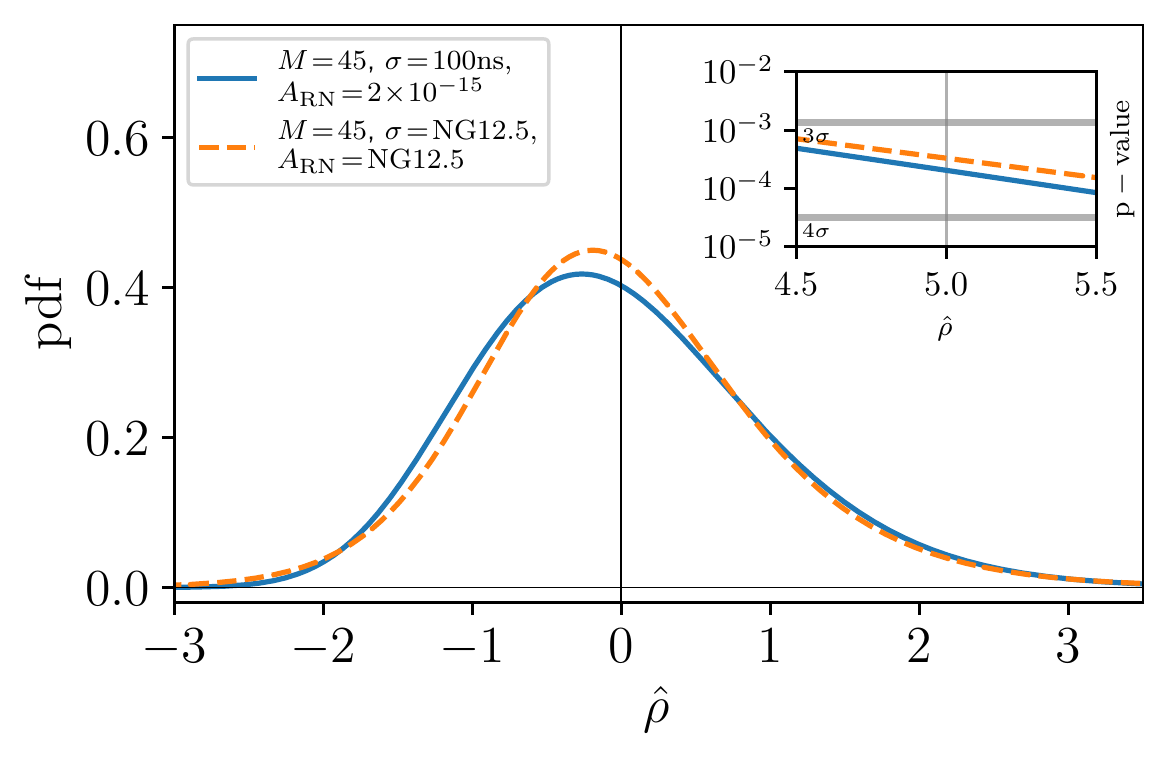}}
\caption{Comparison of GX2 distributions when varying different PTA parameters:
(a) varying the number of pulsars for fixed pulsar white noise;
(b) varying the amplitude of the common-spectrum red-noise process for fixed pulsar white-noise and
number of pulsars;
(c) comparing distributions with different spatial correlations for fixed pulsar white noise and number of pulsars;
(d) comparing distributions having a non-trivial and trivial (i.e., identity) timing model; 
(e) varying the spectral index of the GWB search for fixed white noise and number of pulsars;
(f) comparing a set of realistic NANOGrav parameters, with a simple white-noise plus common-spectrum
red-noise process.
The insets show plots of the $p$-values of the various GX2 distributions 
as functions of $\hat\rho$. The gray bands show the $p$-values for traditional $3\sigma$ and $4\sigma$ detection significances 
based on a unit (standard normal) Gaussian distribution. 
Note that the 11, 22 and 45 pulsar cases all fall along the same $p$-values in (a).}
\label{f:gx2_comparisons}
\end{figure}

\myfig{f:gx2_comparisons} shows various GX2 distributions constructed from
the parameters defining some simple (fictitious) PTAs, with the spectral
properties of the pulsars calculated using the pulsar spectral characterization 
software \texttt{hasasia} \cite{hasasia}.
The panels demonstrate how differences in the GW signal
(spatial correlations, spectral index), noise, and timing parameters change the shape of the
distribution. Note that we have not included a comparison of different WN
levels because the noise power spectral densities cancel out of the 
numerator and denominator of $\tilde{\mb{Q}}$, which defines the GX2 
distribution\footnote{See \eqref{e:Qtilde}, \eqref{e:Qtilde_OS}, 
and \eqref{e:Sigma_OS}, noting
that $\bs{\Sigma}$ is block diagonal for the null distribution case.}.

The interplay of the
parameters makes it difficult to predict exactly what the distribution will
look like.  However the trends for any individual parameter follow the simple
rule that as the sensitivity of the PTA is increased, the tail of the 
GX2 distribution becomes smaller at larger values of $\hat\rho$. 
A few other general observations are:
(i) adding sufficiently large red noise to the pulsars quickly obscures 
the subtle differences in the GX2 distributions, as seen in 
panel (b) of \myfig{f:gx2_comparisons}, and 
(ii) the sky positions of the pulsars matter via the $\tilde{\mb S}_{ab}$ terms
in the expression for $\hat\rho$.

\begin{figure}[h!tbp]
\centering
\includegraphics[width=0.49\textwidth]{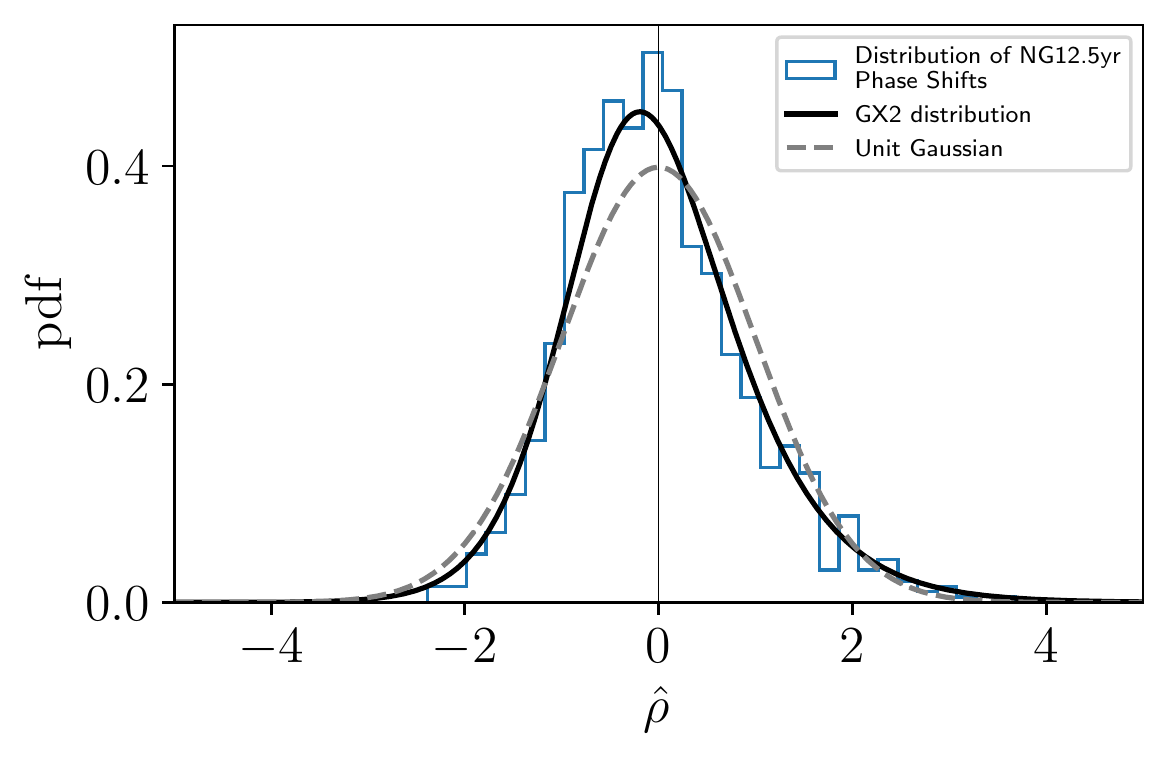}
\includegraphics[width=0.50\textwidth]{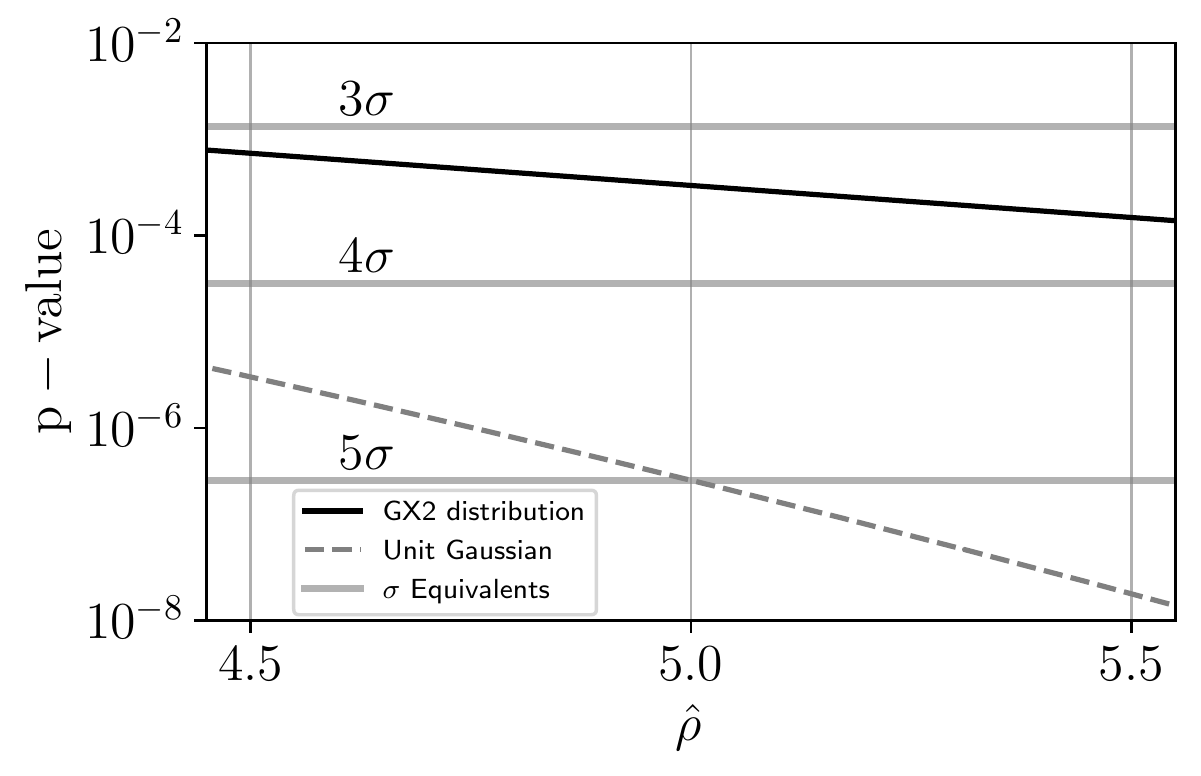}
\caption{Comparison of various distributions of the OS 
signal-to-noise ratio $\hat\rho$ in the absence of 
GW-induced spatial correlations.
Shown are: (i) 
an empirical null distribution for the NANOGrav 12.5-yr data,
obtained from performing 1000 phase shifts (blue histogram),
(ii) the analytic GX2 distribution (black solid line),
and (iii) best-fit (standard normal) 
Gaussian distribution (gray dashed line).
The right panel is a plot of the $p$\,-value as a function of $\hat\rho$.
The phase shifts don't show up in the right-panel plot out to these 
large values of $\hat\rho$.}
\label{f:gx2_gaussian_comparison_zoom}
\end{figure}

Next, we present a realistic GX2 distribution calculated using the noise and
sensitivity parameters of the NANOGrav 12.5-year dataset~\cite{alam:2020nb,
NG12p5yr-stochastic} to demonstrate the usefulness of an accurate analytic 
GX2 distribution in calculating $p$-values or false-alarm probabilities. Diagonalizing the
various matrices described in Sec.~\ref{s:application} over full 
PTA datasets is challenging because of the length of the datasets.
Nonetheless, by using the salient noise and sensitivity parameters 
of the NANOGrav 12.5 dataset, we obtain fairly reasonable agreement 
with the phase-shift method for determining the null distribution
of the optimal statistic.
This is illustrated in \myfig{f:gx2_gaussian_comparison_zoom},
which compares the analytic GX2 distribution with a histogram of 
$\hat\rho$ values for 1000 different 
phase shifts of the NANOGrav 12.5-year data.
For reference, we also show the unit (standard normal) Gaussian distribution.

By looking at the right-hand panel of 
\myfig{f:gx2_gaussian_comparison_zoom}, one immediately sees 
the inaccuracy that would arise if the $p$-value was calculated
assuming that the null distribution was Gaussian.
\mytab{t:p_values} gives the $p$-values for $\hat\rho=5$ and 
$\hat\rho=1.3$ calculated using the analytic form of the 
GX2 distribution, the Gaussian distribution, and from phase-shifts
and sky scrambles of the NANOGrav 12.5-yr data, the latter as
described in \cite{Taylor-et-al:2016b,Cornish-Sampson:2016, abb+2020}.
(The value $\hat\rho=1.3$ is what was measured in the NANOGrav 
12.5-year dataset.)
While the agreement between the various methods is reasonable 
for the lower value of $\hat\rho$, assuming that the null 
distribution is Gaussian for $\hat\rho=5$ leads to a $p$-value 
that is a more than 1000 times smaller than it should be.
\begin{table}[h!tp] 
\begin{center} 
\begin{tabular}{|c|c|c|c|c|} 
\hline
$\quad\hat{\rho}\quad$ & Analytic GX2 & Gaussian & Phase Shifts & Sky Scrambles \\ 
\hline 
5 & $3.3\times10^{-4}$ & $2.87\times10^{-7}$ & \textemdash & \textemdash \\ 
\hline 
1.3 & $0.0983$ & $0.0951$ & 0.091 \cite{abb+2020}& 0.082 \cite{abb+2020}\\
\hline 
\end{tabular} 
\end{center}
\caption{$p$\,-values calculated using various methods in the context of the
NANOGrav 12.5-year dataset.} 
\label{t:p_values} 
\end{table}
%
\section{Discussion}
\label{s:discussion}

We have demonstrated that a generalized chi-squared (GX2) distribution 
is the correct analytical distribution for the optimal 
cross-correlation statistic (OS) used for analyzing PTA datasets.
Although we focused on the null distribution of the OS 
signal-to-noise ratio $\hat\rho$ for this paper, our analyses
in Secs.~\ref{s:rhohat_Agw2} and \ref{s:rho_ab} 
were sufficiently general to show
that GX2 distributions also apply to the optimal estimator of 
the squared-amplitude $\hat A_{\rm gw}^2$ and pulsar-pair 
cross-correlations $\hat\rho_{ab}$ in the presence of a signal.

We applied the general formalism to calculate the GX2 distribution
for $\hat\rho$ using parameters appropriate for 
NANOGrav's 12.5-yr dataset, and 
showed that it agreed quite well with the empirical null 
distribution that
was obtained by phase-shifting the NANOGrav data.
We also calculated the GX2 distribution for several different 
sets of simulated data, in the absence of a GWB cross-correlation,
varying in turn the number of pulsars, the relative contribution 
of red and white noise, etc., to see how these affected the shape
of the resulting distributions. 

Generically, the GX2 distributions we obtained differed from the
best-fit (standard normal) 
Gaussian distribution by having a mode less than their mean and 
having ``fatter'' tails at high values of the statistic. 
(Both distributions have zero mean and unit variance for the 
null distribution case.)
The fatter tails are especially important when calculating $p$-values for 
the null distribution, which is needed to assess the statistical 
significance of a possible detection.

As mentioned in Sec.~\ref{s:results}, constructing the quadratic form 
for the GX2 distribution---which requires solving for the eigenvectors 
of large matrices---is challenging for realistic data. 
Using the full NANOGrav 12.5-year dataset would require solving
the eigenvalue problem for an $N=410064$ dimensional matrix twice. 
An additional complication is that the detailed 
dispersion-measure-variation (DMX) model that NANOGrav uses 
\cite{2015ApJ...813...65T} 
has a large effect on the transmission function of the pulsars 
\cite{Hazboun-et-al:2019,demorest_phd2007}.
This shows up in the timing model fit, which enters the 
quadratic form via the $G$-matrix, e.g.,
$\mb{r}_a = \mb{G}_a^T \bs{\delta}\mb{t}_a$.

As such, the key to constructing a valid GX2 distribution
for a realistic PTA dataset is to find a reduced set 
of parameters that faithfully describes the spectral
properties of the PTA pulsars and the corresponding 
timing model of the array. We used \texttt{hasasia} to calculate the
spectra for all of the NG12.5 pulsars, using their full datasets \cite{Hazboun-et-al:2019}. The noise power
spectral density (from the spectrum) and red noise parameters (from the Bayesian noise analyses) were then used to construct shorter
datasets with similar properties, including the broadband effect of the DMX model. Spectra of these simulated datasets were taken to iteratively
find the correct effective level of the white noise power spectral density to inject into the pulsars
to match the spectra from the full datasets. 

This process proved expeditious to obtain a fairly 
accurate realization of a GX2 distribution on a modest laptop.
But it is computationally inefficient for generating 
GX2 distributions for several different choices of 
noise parameters, for example. Fortunately,
PTA calculations are usually carried out using a rank-reduced 
formalism \cite{vanHaasteren-Vallisneri:2015} that drastically reduces 
the size of matrices dealt with in the analysis. 
This reduced representation of the data alleviates the main 
problem discussed previously.
For example, we are currently developing techniques to use a frequency-domain
implementation of the optimal statistic to speed up the calculation, taking advantage of the
rank-reduced matrix, which is only $2N_{\rm freq}\times 2N_{\rm freq}$,
as opposed to $N_{\rm TOA}\times N_{\rm TOA}$.
But we leave that discussion for future work. 

An alternative to using noise estimates from a previous
Bayesian inference run to calculate the optimal statistic 
is to marginalize over the noise parameters.
As described in \cite{Vigeland-et-al:2018}, marginalizing 
over the red noise parameters tends to remove biases that 
would otherwise exist due to correlations between the 
noise estimates and the timing residual data 
used to construct the optimal statistic.
For the analyses described in this paper, we 
used maximum likelihood estimates of noise parameters 
from a Bayesian analysis to construct the quadratic forms 
needed for calculating the GX2 distributions. 
We did not investigate any source of bias that might 
have been introduced by using noise estimates as
opposed to noise marginalization.
But we are currently investigating the possibility of 
marginalizing over the noise for future uses of the
GX2 distributions.

Finally, optimal cross-correlation statistics are also used when 
analyzing data from ground-based GW detectors like Advanced 
LIGO, Virgo, and KAGRA.
However, for this case, the optimal statistics are 
well-described by Gaussian distributions, so GX2 
distributions are not needed. 
This is because the data from pairs of detectors are analyzed
in roughly 100-sec segments (to account for potential
non-stationarities in the detector noise power) and then
averaged together over $\gtrsim 10^5$ such segments,
corresponding to a typical year-long observation.
The cross-correlation estimates of the amplitude of the 
GWB for each 100-sec segement are GX2 distributed.
But the final averaged optimal-statistic value (inverse-noise-weighted 
by the variance of the individual estimates) is 
well-described by a Gaussian distribution due to the 
central-limit theorem.

\section*{Acknowledgements}
\label{s:acknowledgements}

JSH, PMM, JDR and XS acknowledge support from the NSF NANOGrav Physics Frontier Center 
(NSF grants PHY-1430284 and PHY-2020265).
JDR acknowledges support from start-up funds from Texas Tech University. JSH
acknowledges support from start-up funds from Oregon State University. The
authors thank \texttt{Math.StackExchange} user River Li, whose
answer\footnote{\url{https://math.stackexchange.com/a/3716276}} introduced us
to the technique of diagonalization to derive probability distributions for
inner products. Lastly, we also thank the authors of \cite{dg2021} for helpful
discussions regarding the MatLab
code\footnote{\url{https://www.mathworks.com/matlabcentral/fileexchange/85028-generalized-chi-square-distribution}}
they developed to calculate GX2 distributions. 

\bibliography{refs}

\appendix

\section{Empirical tail fitting and extrapolation}
\label{s:empirical_tail_fitting}

The GX2 is the analytic null distribution for the optimal statistic, but
it depends upon specific characteristics of the dataset e.g., white noise and red noise parameters.
In the case of red noise, a Bayesian analysis is used to obtain a posterior distribution on the amplitude and spectral index,
and using different draws from that posterior to calculate the GX2 changes the resulting distribution.
Choosing the best parameters to use requires care and potentially comparison with empirical distributions.
In this appendix we introduce a technique that can be used in the interim.
In situations where we need to evaluate the null distribution in a place where we have few
empirical simulations, but we don't yet have the full GX2 method, we can empirically
fit the tail of the empirical simulations with an exponential function.
This method has been used to extrapolate the tail of the null distribution in e.g., searches for
GWs from rapidly rotating neutron stars~\cite{AbbottAbbott2022}.
The uncertainty on the fit parameters can then be used to bound our estimate of the 
$p$-value or false alarm probability.

We begin by choosing a point beyond which $p(\hat\rho)$ looks like an exponential.
This choice is arbitrary, but the procedure can be done using several choices, picking the
one that looks most reasonable. We call this value $\hat\rho_{\mathrm{tail}}$.
Therefore, for $\hat\rho > \hat\rho_{\mathrm{tail}}$, we have
\be
    p(\hat\rho | \hat\rho > \hat\rho_{\mathrm{tail}}) = \lambda \exp\left[-\lambda \left(\hat\rho - \hat\rho_{\mathrm{tail}}\right)\right]\,.
\ee
The shape parameter $\lambda$ is a free parameter that we fit
using the values in the tail of the empirical distribution.
Since there is only a single free parameter, we can fit this using a brute-force Bayesian approach,
with a uniform prior on $\lambda$
\begin{align}
    p[\lambda | \left\{\hat\rho\right\}_{i=1}^{N_{\mathrm{tail}}}] = \frac{1}{\lambda_{\mathrm{max}}}\lambda^{N_\mathrm{tail}} \exp{\left[-\lambda \sum_{i=1}^{N_\mathrm{tail}}\left(\hat\rho_{i} - \hat\rho_{\mathrm{tail}}\right)\right]}\,,
\end{align}
where $N_{\mathrm{tail}}$ is the number of empirical distribution values satisfying $\hat\rho_{i} > \hat\rho_{\mathrm{tail}}$,
and $\lambda_{\mathrm{max}}$ is the upper bound on the uniform prior on $\lambda$.
In this notation, $\{\hat\rho_i\}_{i=1}^{N_{\textrm{tail}}}$ denotes the set of empirical distribution results that populate the tail.

\begin{figure}
    \includegraphics*[width=0.49\textwidth]{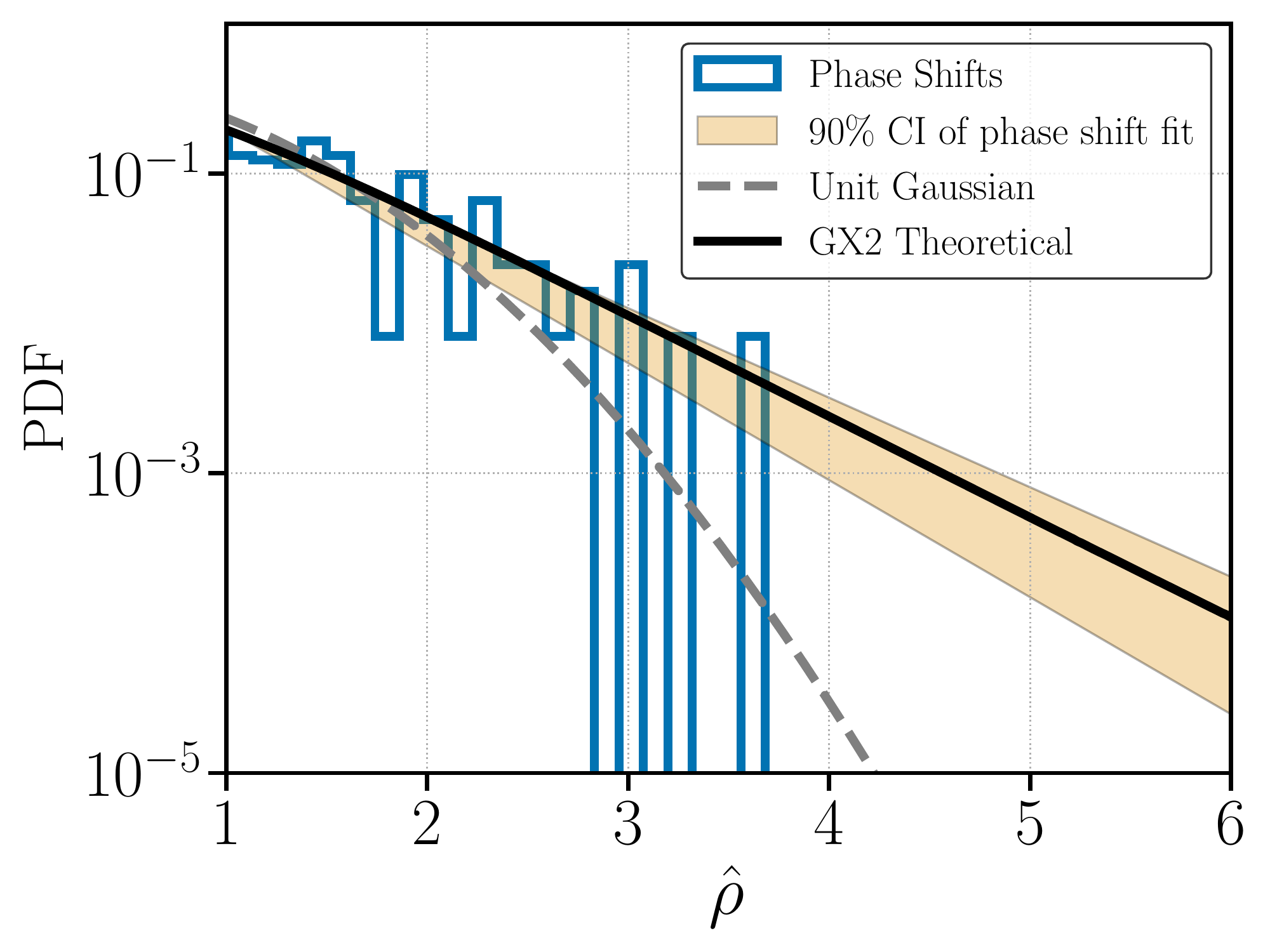}
    \includegraphics*[width=0.49\textwidth]{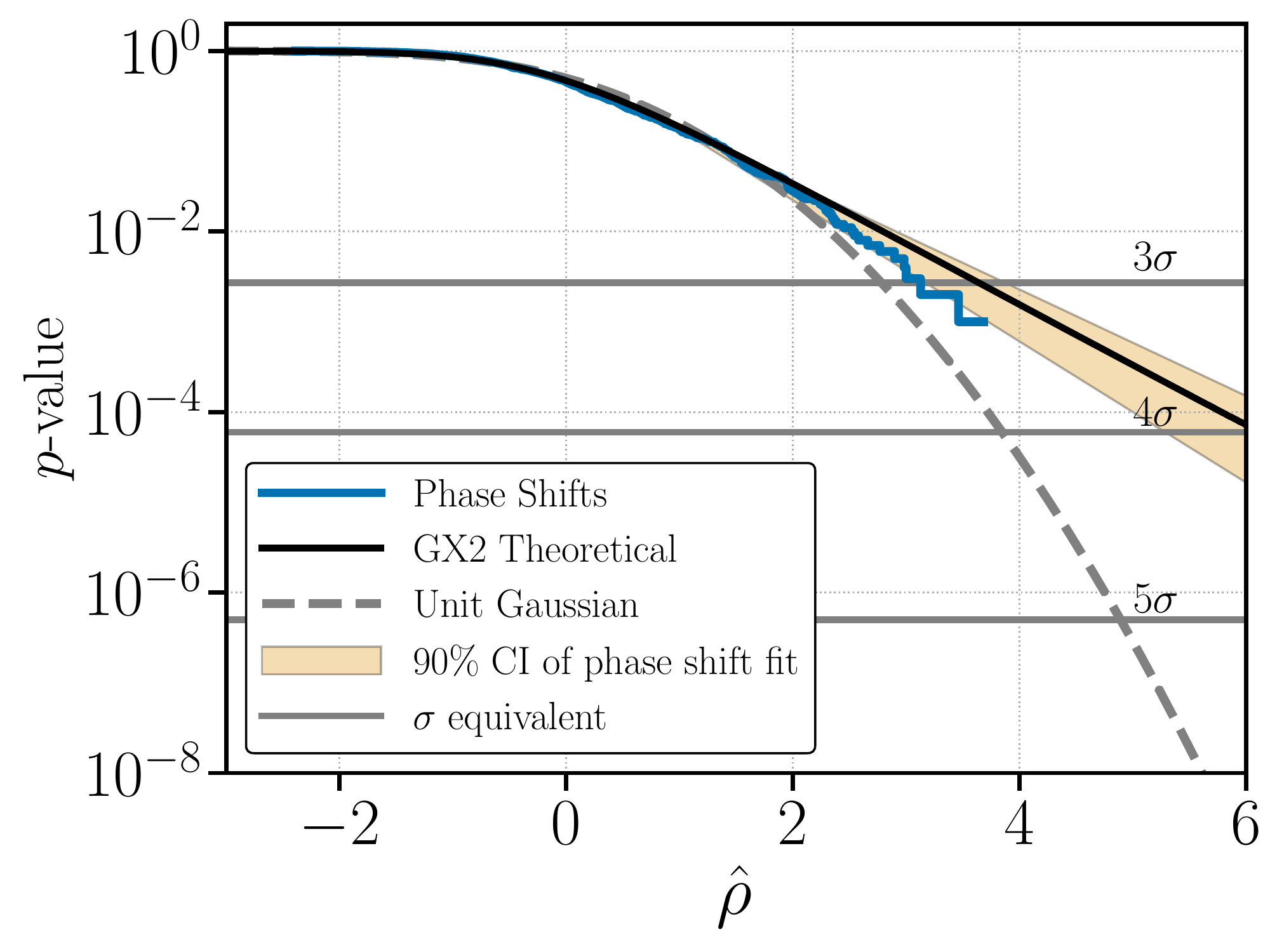}
    \caption{Comparison of GX2 fit to empirical phase shifts and  tail fitting procedure. Left: 90\% credible interval of fit (orange) to the empirical PDF (blue). The theoretical distribution from the GX2 distribution is shown in black. Right: The same but for the $p$-value $(1 - {\rm CDF})$. 
In both cases we can see that fitting the empirical distribution from phase shifts with an exponential gives a reasonable approximation of the GX2 in the region in which we are interested.}
    \label{fig:tail_fitting_example}
\end{figure}

The posterior on $\lambda$  can then be used to quantify the uncertainty in the tail
of the distribution on $\hat\rho$.
We show an example of fits to the tail of the empirical distribution in \myfig{fig:tail_fitting_example} 
using 300 different draws from the posterior on $\lambda$ 
(the 90\% credible interval is shown in orange), along with
the GX2 estimate (black line), and the empirical distribution (blue histogram).
The blue histogram uses the phase shifts from the NANOGrav 12.5-year analysis that were used to generate \myfig{f:gx2_gaussian_comparison_zoom}.
For the sake of comparison we also include the best-fit Gaussian PDF (grey dashed), which clearly underestimates $p(\hat\rho)$ in the tail.
The exponential fit agrees with the GX2 estimate out to $\hat\rho=6$,
indicating it can be effectively used to extrapolate the tail and still be consistent with the analytic distribution.
We can also quantify the uncertainty in the tails of the empirical distribution due to having a small number of points in the tail of the empirical distribution.

Once we have a fit for $p(\hat\rho)$,
we can analytically calculate the cumulative distribution function (CDF), $p(\hat\rho < \rho)$, which can be used to estimate the false alarm probability.
The CDF is given by
\begin{align}
    p(\hat\rho < \rho|\lambda) = \begin{cases} 
    \frac{1}{N}\sum_{i=1}^N \mathbbm{1}_{\hat\rho_i \leq \rho} & \rho \leq \hat\rho_{\mathrm{tail}}\,,\\[2em]
    \frac{1}{N}\sum_{i=1}^N \mathbbm{1}_{\hat\rho_i \leq \hat\rho_{\mathrm{tail}}} + \mathcal N\left(1 - e^{-\lambda (\rho - \hat\rho_\mathrm{tail})}\right)& \rho > \hat\rho_{\mathrm{tail}}\,,
    \end{cases}
 \end{align}
 where $\mathbbm 1$ is the indicator function that equals 1 if the subscript condition is met and equals zero otherwise.
 In this expression, $N$ is the total number of empirical trials and $\hat\rho_{i}$ is now any one of those $N$ trials 
(as opposed to being taken only from the tail).
 The normalization factor $\mathcal N$ is chosen such that the CDF evaluates to 1 at infinity.

\end{document}